\documentstyle[12pt]{article}  
\date{ }

\title{Spacetime models, fundamental interactions and noncommutative geometry}
\author{J. S\l adkowski* \\  Institute of Physics, 
University of Silesia, \\ Uniwersytecka 4, Pl 40007 Katowice,\\  Poland}

\begin{document}
\baselineskip9mm
\maketitle
\begin{abstract}
\baselineskip7mm
We discuss the problem of determining the spacetime structure. 
We show that when we are using only topological methods the spacetime can be 
modelled as an ${\bf R}$- or ${\bf Q}$-compact space although the 
${\bf R}$-compact spaces seem to be more appropriate. Demanding 
the existence of a differential structure substantially narrows the 
choice of possible models. The determination of the differential 
structure may be difficult if it is not unique. By using the noncommutative 
geometry construction of the standard model we show that fundamental 
interactions determine the spacetime in the class of ${\bf R}$-compact 
spaces. Fermions are essential for the process of determining the 
spacetime structure.

\end{abstract}
\vspace{25mm}
* e-mail: sladk@us.edu.pl
\newpage
\section{Introduction}

\ \ \ The outcomes of physical measurements are expressed in rational 
numbers. We believe that 
all possible values of physical variables constitute the set of real numbers 
{\bf R}. It is an idealized view since all measurements are performed with 
certain accuracy and it is hard to imagine how can they give irrational 
numbers. In this way the algebra of real continuous functions $C(M)$ on 
the spacetime manifold $M$ comes to play. This algebra play central r\^ ole in 
classical and quantum physics, although this fact is not always perceived. 
Most of physical theories, including quantum gravity, make use of the notion 
of spacetime, at least approximately. Therefore  
one of the most important and fundamental open problems in theoretical 
physics is to explain the origin and structure of spacetime. Here we would like 
to discuss the problem of determining the spacetime structure. Put it 
another way, to analyse how faithful our theoretical models 
of the spacetime can be. 
We will try to be model independent and avoid unnecessary 
assumptions. Nevertheless, we will suppose that it is possible to determine 
the algebra $C(M)$ on the spacetime (assumed to be a topological space) with 
sufficient for our aim accuracy. This does not mean that we have to be able to 
find each element of $C(M)$ "experimentally": some inductive construction 
should be sufficient. By 
an abuse of language, we will call elements of $C(M)$ observables. 
Then we will discuss to what extent the structure of the model $M$ of 
the spacetime is determined by $C(M)$, $M$ being a topological space. Further, 
we will analyse what happens if we admit of $M$ to have no topology
or to be a differential manifold. We will also use the algebra of continuous 
$K$-valued functions $C(M,K)$, $K$ being a topological ring. Finally, we will 
show how $C(M,K)$ can be used to construct field theory via the A. Connes 
construction. We will also discuss to what extent the spacetime manifold 
is determined by electroweak interactions in the Connes' noncommutative 
geometry formalism.
 
\section{In quest of the topology of spacetime}

\ \ \ A lot of properties of a topological space $M$ is encoded in the 
associated algebras $C(M,K)$ of continuous $K$-valued functions, $K$ being 
a topological ring, field, algebra etc. Even  differential structures on a 
manifold $M$ can be equivalently defined by appropriate subalgebras
$C^{k}(M,K)$ of real or complex differentiable functions on $M$. Suppose that 
our experimental technique is a priori powerful enough to reconstruct 
$C(M,{\bf R})\equiv C(M)$ on our model of the spacetime $M$. What sort of 
information concerning $M$ can be extracted from these data? If $M$ is a set 
and ${\cal C}$ a family of real functions $M \rightarrow {\bf R}$ then 
${\cal C}$ determines a (minimal) topology $\tau _{\cal C}$ on $M$ such that 
all function in ${\cal C}$  are continuous [1-2]. In general, there will be
real continuous functions on $M$ that do not belong to ${\cal C}$ and 
more families of real functions on $M$ would define the same topology on $M$. 
$(M, \tau _{\cal C})$ is a 
Hausdorff space if and only if for every pair of different points $p_{1},p_{2} 
\in M$ there is a function $f \in {\cal C}$ such that $f(p_{1})\not= 
f(p_{2})$. Therefore it seems reasonable to assume that 

$$f\left( x\right) = f\left( y\right)\ \ \forall f\in C\left( M
\right)\ \ \ \Rightarrow \ \ \  x=y .\eqno(*)$$
Physically this means that  in order to be able to distinguish $x$ 
from $y$ in our model of spacetime we have to find such an 
observable $f\in C(M)$ 
that for $x,y\in M \  \ f(x) \not= f(y)$. From the mathematical point of view, 
we have to identify all points that are not distinguished by $C(M)$, that is 
to demand (*). It is easy to show that such spaces are Hausdorff spaces. To 
proceed let us define [2-4]. \\ 

{\bf Definition 1.} \ \ \ Let $E$ be a topological space. A topological 
Hausdorff space $X$ is called 
$E$-compact ($E$-regular) if it is homeomorphic to a closed (arbitrary) 
subspace of some Tychonoff power of $E$, $E^{Y}$. \\ 

The following facts justify our assumption (*). For a topological space 
$X$, not necessarily a Hausdorff one, we can construct an $E$-regular space 
$\tau _{E}X$ and its $E$-compact extension $\upsilon _{E}X$ 
so that we have [3-4] 

$$C\left( X,E\right) \cong C\left( \tau _{E} X,E\right) \cong 
C\left( \upsilon _{E}X,E\right)\cong
C\left( \upsilon _{E} \tau _{E}X,E\right) ,$$ 
where $\cong$ denotes isomorphism. The spaces $\tau _{E}X$ and 
$\upsilon _{E} \tau _{E}X $ have the nice property (*). Now, it is obvious 
that, in general, our theoretical model of the spacetime may not be uniquely 
determined. This is an important 
result that says we can always model our spacetime as a subset of some 
Tychonoff power of ${\bf R}$ provided $C(M)$ is known! But it also says that 
we can model it as a subset of a Tychonoff power of a different topological 
space e.g. the rational numbers ${\bf Q}$ (cf the discussion at the 
beginning). So its our choice! The topological number fields  ${\bf R}$ 
and ${\bf Q}$ have the additional nice property of determining uniquely 
(up to a homeomorphism) ${\bf R}$- and ${\bf Q}$-compact sets, respectively:

$$C\left( X,E\right) \cong C\left( Y,E\right) \iff X\ is \ 
homeomorphic\ to\ Y,\ E={\bf R}\ or\ {\bf Q}. \eqno(**) $$
Other topological rings can also have this property. 
But this does not mean that the spacetime modeled by $C(M,E)$ is homeomorphic 
to the one modeled by $C(M,E')$. Hewitt have shown that ${\bf R}$-compact
spaces are determined up to a homeomorphism by $C(X,E)$, 
where $E={\bf R}$, ${\bf C}$ or ${\bf H}$, the topological fields of complex 
numbers and quaternions, respectively [5]. 
This means that if we are interested in 
modeling spacetime as an ${\bf R}$-compact (${\bf Q}$-compact) space then 
we can use $C(M,{\bf R}), \ C(M,{\bf C})$ or $C(M,{\bf H})$ 
($C(M,{\bf Q})$) to determine it. Another problem we will face is to 
decide if we are dealing with the algebra $C(X,E)$ or only with the algebra 
of all continuous bounded $E$-valued functions on $X$, 
$C^{*}(X,E)$ [2-4]. For a compact space $X$ we have $C(X,E)=C^{*}(X,E)$, 
but in general, they are distinct. Spaces on which 
all continuous real functions 
are bounded are called pseudocompact. An ${\bf R}$-compact pseudocompact 
space is compact. We might get hints that 
some observables may in fact be unbounded but we are unlikely to be able to 
"measure infinities". An unbounded observable is necessary to show that the 
spacetime is a noncompact topological space. If we suppose that we can 
only recover $C^{*}(M,{\bf R})\equiv C^{*}(M)$, then we 
can as well suppose that $M$ is compact (for an ${\bf R}$-compact $M$). 
In general, there will be more spaces with $C^{*}(M)$ as the algebra of 
real bounded continuous functions on them (they may not be compact 
or even ${\bf R}$-compact). Compactness (or paracompactness) of the space 
is a welcome property. For example pseudodifferential operators have 
discrete spectrum on compact spaces. Physicists often compactify 
configuration spaces by adding extra points or imposing appropriate 
boundary conditions.  Demanding that all physical fields vanish at 
infinity is usually equivalent to the one point compactification of the 
spacetime  and requiring that all fields vanish at the added "infinity 
point". In general, a topological space $X$ has more then 
one compactification. In some sense the one point compactification is 
minimal and the Stone-\^ Cech compactification is maximal [2]. We will 
probably have to make nontopological assumptions to choose one 
among the possible compactifications although they can be 
distinguished by regular subrings of $C(M)$ if they contain constant 
functions [3-4]. \\ 

{\bf Definition 2.} \ \ \ We shall say that a subspace $X$ of $M$ is 
$C$-embedded in $M$ if every function in $C(X,E)$ can be extended to a function 
in $C(M,E)$. Likewise, we shall say that $X$ is $C^{*}$-embedded in $M$ if 
every function in $C^{*}(X,E)$ can be extended to a function in 
$C^{*}(M,E)$. \\ 

A priori, after determining $C(M,E)$ or $C^{*}(M,E)$ we may find out that 
some space in which $M$ is $C$- or $C^{*}$-embedded is as good a model of the 
spacetime as $M$ is, and vice versa. Fortunately, for most topological 
spaces $X$ (completely ${\bf R}$-regular ones [2-5]) there is a unique compact 
space $\beta X$ (the Stone-\^ Cech compactification) in which $X$ is 
dense and $C^{*}$-embedded and a unique ${\bf R}$-compact space 
$\upsilon _{\bf R} X$ in which $X$ is dense and $C$-embedded [2]. 
It can be proven that $\upsilon _{\bf R} X$ can be embedded in 
$\beta X$ and that $\upsilon _{\bf R} X$ is the smallest ${\bf R}$-compact 
space between $X$ and $\beta X$ [2-4]. The spaces $\beta X$ and 
$\upsilon _{\bf R} X$ are in some sense (see below) upper 
and lower limits on the spaces 
we are looking for. As $C(X)$ distinguishes among ${\bf R}$-compact 
spaces [2], we have to find a physical phenomenon that is not describable 
in terms of $C(X)$ to prove the assumption that the spacetime is an 
${\bf R}$-compact space to be wrong. In general one can say that $C(X,K)$ is more 
sensitive than $C^{*}(X,K)$ (e.g. it can distinguish between $X$ and $\beta X$). 
The following theorems give us some sense of the limitations of the 
determination of the spacetime modeled by $C(M,{\bf R})$ [2].\\
 
{\bf Theorem 1.} \ \ \  If $X$ is dense in $T$ then the following statements 
are equivalent. \\
{\it i}\ \ \  Every continuous mapping from $X$ into any ${\bf R}$-compact 
space $Y$ has an \ \ \ extension to a continuous mapping from $T$ to $Y$. \\ 
{\it ii}\ \ \ $ X \ \subset  \ T \ \subset \upsilon _{\bf R} X$. \\
{\it iii}\  \ $\upsilon _{\bf R} T = \upsilon _{\bf R} X$. \\
{\it iiii}\ $X$ is $C$-embedded in $T$.

{\bf Theorem 2.} \ $\upsilon  _{\bf R}Y$ contains a $C$-embedded copy of $X$ if 
and only if $C(X,{\bf R})$ is a homeomorphic image of  $C(Y,{\bf R})$. \\ 

{\bf Theorem 3.} \ \ If $X$ is dense in $T$ then the following statements are 
equivalent. \\
{\it i}\ \ \  $X$ is $C^{*}$-embedded in $T$. \\ 
{\it ii}\ \ \  $X \ \subset  T \subset \beta X$. \\ 
{\it iii}\ \ \  $\beta T \ =\ \beta X$.

{\bf Theorem 4.}  $\beta Y$ contains a $C^{*}$-embedded copy of $X$ if 
and only if $C^{*}(X,{\bf R})$ is a homeomorphic image of  
$C^{*}(Y,{\bf R})$. \\ 

One can try to estimate the cardinality of the difference between various 
spaces in question. Theorems 5 and 6 [2] say that it can be essential. \\ 

{\bf Theorem 5.} If $X$ is locally compact and ${\bf R}$-compact then the 
cardinal of a closed infinite set in $\beta X - X$ is at least $2^{\bf c}$. \\ 

{\bf Theorem 6.} The cardinal of a nondiscrete, closed set in 
$\beta X -\upsilon _{\bf R}X$ is at least $2^{\bf c}$. \\ 

Physicists frequently raise questions concerning the potential discreteness 
of spacetime. One can  formulate conditions of finiteness in terms of $C(X,K)$ 
[2-4]. It is unlikely that the spacetime forms a finite set (although various 
finite approximation have been put forward [6]). The answer to the question 
if discreteness can be defined in terms of $C(X,K)$ depends on the axioms of 
set theory! If one assume the existence of measurable cardinals, then 
conditions of discreteness of $X$ cannot be formulated in terms of 
$C(X,K)$ or $C^{*}(X,K)$, [3-4]. Nevertheless, the following theorem can be 
proven [2]: \\ 

{\bf Theorem 7.} \ \ \ A discrete space is ${\bf R}$-compact if and only 
if its cardinal is nonmeasurable. \\

The existence of measurable cardinals cannot be proven in the standard 
axioms of set theory. Even if they do exist they must 
be so huge that it is unlikely that the spacetime is so "potent". Therefore 
if the spacetime is discrete we certainly will be able to model it as 
an ${\bf R}$-compact space and discover this fact "on inspection" of $C(M,E)$, 
$E={\bf R,C,H}$. Cardinality of such space can also be enormous (e.g. ${\bf c}$, 
$2^{\bf c}$, $2^{2^{\bf c}}$, ...). \\ 

\ \ \  The problems of cardinality, dimension, density and tightness 
of the spacetime can also be addressed in terms of rings of real continuous 
functions with various topologies although experimental verification 
of these features (except dimension) 
is unlikely. The reader is referred to [7] for details. Here, we would like 
to mention only the following two facts. 
${\bf R}$-compact spaces $X$ are precisely those with countable Hewitt 
numbers, $q(X) \le \aleph _{0}$ [7]. For an arbitrary topological space and 
cardinal $\tau $ there is a subspace $\nu _{\tau} X$ of $\beta X$ so that 
every continuous function $f: X\rightarrow {\bf R}$ can be extended to a 
continuous real function on $\nu _{\tau} X$ and $q(\nu _{\tau} X) 
\le \tau $ [7]. \\ 

It may be too optimistic to assume that we are able 
to determine $C(M,{\bf R})$ 
with the required precision. Suppose that our experimental technique 
allows only for sort of {\it yes} or {\it no} answer to questions 
concerning spacetime structure [8]. In this case we have to consider 
determination of a topological space $X$ by the ring $C(X, D)$ of 
continuous functions into $D=\{ 0, 1\} $ with various topological 
and/or algebraic structures. In general, $C(X,D)$ does not determine the 
space $X$ although $C(X,{\bf Z}_{2})$ fulfils (**) with $E={\bf Z}_{2}$. 
One can also consider other discrete fields e.g. ${\bf Z}_{3}$ [3-4]. 
In such case we can only try to determine the space in the class of 
$E$-compact spaces for some discrete $E$. Topological subfields of ${\bf R}$ 
can also be used for that purpose because they fulfil (**) [2,3,9]. \\ 

Most of physical models of spacetime require that it is metrizable. 
Metrizable spaces with nonmeasurable cardinals are ${\bf R}$-compact [2]. 
This means that "practically all" models of spacetime are  ${\bf R}$-compact 
(cf the discussion of discreteness). \\ 

Up to now we have considered the arbitrariness of our mathematical 
model $X$ of the spacetime as determined by $C(X,{\bf R})$. But one can also 
ask if any algebra that we identify as an algebra of physical observables 
on the spacetime always defines a topological space. The answer is negative: 
a commutative algebra must fulfil various sets of conditions to be a $C(X,{\bf 
R})$ of some topological space $X$. If we suppose that our model of the 
spacetime is not a topological space we can deal with ${\bf R}^{X}$, the 
algebra of all real functions on $X$. But to have some "deterministic power" 
we have to demand the existence of some additional structure on 
$X$, that is to distinguish 
a family of subsets of $X$ and/or an algebraic structure on the class of 
functions we are dealing with [10]. For example, 
if $(X,\tau )$ is a pair consisting of 
a set $X$ and a family $\tau $ of its subsets then we can define "continuity" 
and "homeomorphisms" by replacing topology by the family $\tau $. In this case 
one can prove [3-4]. \\ 

{\bf Theorem 8.} \ \ \ Let $X$ and $Y$ be sets and $\tau $ and $\sigma $ 
families of their subsets containing the empty set, closed with respect to  
finite intersections and summing up to $X$ and $Y$, respectively. Then 
$X$ and $Y$ are "homeomorphic" if and only if there is an isomorphism 
of the semigroups $D^{X}$ and $D^{Y}$ such that $"C(X,D)"$ is mapped onto 
$"C(Y,D)"$. \\ 

Such generalized space are more difficult to deal with than ordinary 
topological spaces therefore we think that spacetime should be modelled 
in the class of topological spaces.

One may also wonder if the knowledge of some symmetries might be of any help. 
In general, a topological space $X$ is not determined by its symmetries 
(homeomorphisms $X \rightarrow X$) [12-13] but 
sometimes can provide us with useful information, e. g.  
if we know that some group $G$ acts transitively on $X$ then the cardinality 
of $X$ is not greater than the cardinality of $G$ [14]. For example, 
if we are pretty sure that the Lorentz group acts transitively on the 
spacetime we have got an upper bound on the cardinality of the spacetime. \\ 

Let us sum up the above consideration. We can model the spacetime 
as a topological ${\bf R}$-compact or ${\bf Q}$-compact space 
although the ${\bf R}$-compact spaces seem to be more appropriate. This 
two spaces are not necessarily homeomorphic. We might have serious problems 
with identification of some of the topological properties of the spacetime. 
This is because more then one space will have the same algebra of $C(M,E)$. 
If we decide to model the spacetime as a (completely regular) 
${\bf R}$-compact space $M$ then we are able to reconstruct $M$ from $C(M)$ 
 or $C^{*}(M)$ in the following sense [2,11]. $C(M)$  or $C^{*}(M)$ 
determine its Stone-\^ Cech 
compactification $\beta M$ with $M$ as a dense ${\bf R}$-compact 
subspace. All fixed ideals in 
$C(M)$ correspond to points in $M$ [2-4,11]. Such spaces are Hausdorff. In 
order to distinguish two spacetime points we need an observable that takes 
different values at these points. If we fail to do this we have to identify 
these points and this may result in a discrete or even finite 
model that would also be 
an ${\bf R}$-compact space and can be reconstructed from $C(M)$. Of course, 
spacetime points may have "reach structure" that is beyond our experimental 
scope. This corresponds to determining only some subalgebra of $C(M)$. 
We have to find a phenomenon that is not describable in terms of 
$C(M)$ to reject the assumptions of ${\bf R}$-compactness.  We do 
not know if the physical world can be described by using only topological 
methods. The most spectacular example is the existence 
of the Whitehead spaces. 
These are three-dimensional topological manifolds that are not 
homeomorphic to ${\bf R}^{3}$ 
but their products with ${\bf R}$ are homeomorphic to  ${\bf R}^{4}$. 
In other words when an ${\bf R}^{1}$ is factored out in ${\bf R}^{4}$ 
the result will not necessary be ${\bf R}^{4}$. One have to demand 
differentiability for this to be case. 
More sophisticated formalism would involve further assumptions about 
the spacetime structure but it may not be easy to find out if these 
assumptions are necessary or just convenient tools. We will discuss 
it in the following sections.

\section{Differential structure.}
 
\ \ \ Differential calculus have proven to be a powerful tool in the  hands 
of physicists. But is it indispensable? Not every topological space or even 
topological manifold
can support differential structures and demanding the existence of a 
differential structure on the spacetime can severely restrict our choice of 
spaces for modeling the spacetime. A differential structure on a topological 
manifold $M$ can be defined by specifying a subalgebra of differentiable 
functions  $C^{k}(M,{\bf R})$ of the algebra $C(M)$. The algebra 
$C^{\infty}(M)$ of smooth real 
functions on $M$ determines $M$ up to a diffeomophism [16] (the points of $M$ 
are in one-to-one correspondence with maximal ideals in $C^{\infty}(M))$. 
The algebra of continuous function on $M$ is larger than $C^{k}(M,{\bf R})$ 
and may correspond to more topological spaces than $M$ but if two manifolds 
have at some points $p$ and $q$ isomorphic rings of germs of continuous 
functions then the points $p$ and $q$ have homeomorphic neighbourhoods 
(local dimensions are the same) [17]. If the laws of 
physics are "smooth" the spacetime should be modeled on a smooth manifold. If 
this is the case then $C^{\infty}(M,{\bf R})$ is sufficient to 
determine $M$ and
describe all physical phenomena. Geometrical quantization is one of the most 
popular efforts in this direction. One can even try to "quantize" differential 
equations in terms of $C^{k}(M,{\bf R})$ [18]. It should be noted here that 
$C(M)$ or even $C^{\infty}(M)$ are far to big as potential algebras of 
observables to be used for constructing a quantum theory that agrees with 
experiment. Additional information (assumptions) is necessary to deal with 
this problem [19]. Any manifold can be embedded in ${\bf R}^{n}$ for some $n$ 
and therefore is ${\bf R}$-regular. The most popular models of spacetime 
are riemannian or pseudoriemannian manifolds. 
Such spaces are metrizable and as such 
${\bf R}$-compact (cf the discussion in the previous section). This means that 
these manifolds are as topological spaces determined by $C(X,E)$, where 
$E={\bf R}$, ${\bf C}$ or ${\bf H}$ but additional knowledge 
of the algebra of differentiable functions is needed to determine the 
differential structure [16].
But even in the smooth case 
we face a new nonuniqueness problem because some manifolds can support many 
nonequivalent differential structures [20-27]. Such "additional" differential 
structures are usually referred to as  {\it fake } or  {\it exotic } ones.
They are specially abundant in the fourdimensional case (it is sufficient to 
remove one point from a given manifold to get a manifold with exotic 
structures [24]). More astonishing is the fact 
that the topologically trivial fourdimensional Euclidean space ${\bf R^{4}}$ 
can be given uncountably many exotic structures (in fact a 
two-parameter family of them) [24]. We have to interpret 
these mathematical results in physical language [25-27]. This is not an easy 
task. Although one can put forward many arguments that exotic smoothness might 
have physical sense [26,30-31], the lack of any explicit (pseudo-) riemannian 
structure  hinders physical predictions. Nevertheless some 
problems can be discussed. Suppose that the 
spacetime manifold is topologically ${\bf R}^{4}$. If we require that all 
physical observables vanish at infinity then our model is equivalent to the 
one on ${\bf S}^{4}$ with the "boundary condition" that all observables vanish 
at one point. Then if the smooth Poincare hypothesis (there is only one 
differential structure on ${\bf S}^{4}$) [28] is correct we are left with only 
one (standard) differential structure. This may be a solution to the 
nonuniqueness problem but certainly is not a satisfactory explanation of the 
fact! If we suppose that all observables have compact supports then we are not 
able to eliminate exotic structures [27]. Bizaca and Etnyre proved that for 
any compact 3-manifold $M$ the open manifold $M\times {\bf R}$ has infinitely 
many different smooth structures (this is true for a wider class of $M$'s) 
[29]. Spaces of these form are often explored by physicists. Below we give 
some examples.  Bag models are popular models of hadrons and astrophysical 
objects [32-33].  Such models may have their exotic versions because we 
do not know if the choice the of metric tensor and boundary conditions is 
sufficient to eliminate the possible exotic structures, especially in 
astrophysical considerations. We are accustomed to the coordinate 
representation of quantum mechanics. Most of the configuration spaces for 
quantum problems are of the form $M\times {\bf R}$. If the Schr\"odinger 
operator is of the form  $-\triangle + V$, where $\triangle $ is the 
Laplace operator and $V$ the potential then some $\triangle $ may not be
consistent with all smooth structures (that is the metric tensor may not 
be smooth). We think that if one chooses the metric tensor and boundary 
condition for the above problems then the additional differential 
structures are physically unimportant. This means that we are only interested 
in the spectral problem of the operators in 
question and suppose that physically 
interesting isospectral homeomorphic manifolds are diffeomorphic, cf [30-31]. 
But in general relativity metric tensor is one of the variables 
and the question is what 
determines differential structure in this case? Once more we dare to 
conjecture that the spectral problem for physical operators should give an  
answer to this nonuniqueness problem. Unfortunately, our present knowledge 
is too poor to give a definite answer. H. Brans has conjectured that 
"localized" exoticness can act as a source for some externally 
regular gravitational field, just as matter or a wormhole can [25].
In this context one can also ask if there is an analogue of the 
Bohm-Aharonov effect. That is suppose  that some points are "excluded" from 
the spacetime. Such singularities allows for exotic structures. The "standard" 
metric tensor defined  by matter might not be smooth with 
respect to some exotic 
differential structures. Can such effect be detected, say in gravitational 
measurements? This would mean that there is 
"additional" curvature  required by consistency of differential structures. 
The existence of exotic differential structures is certainly a challenge to  
physicists [26]. We will return to this problem in the following section. 

\section{Noncommutative differential geometry and physical models.}

\ \ \ We have seen that differential geometry can be formulated in terms of 
the commutative algebra of real smooth functions on the manifold in question. 
A. Connes managed to generalize it for much larger class of algebras, not 
necessarily commutative [34-35]. His noncommutative geometry have found 
profound physical applications. The basic ingredients are a $C^{*}$-algebra 
${\cal A}$ represented in some Hilbert space ${\it H}$ and an operator 
${\cal D}$ acting in ${\it H}$. The differential $da$ of an $a \in {\cal A}$ 
is defined by $[{\cal D},a]$ and the integral is replaced by the Diximier 
trace, $Tr_{\omega}$, with an appropriate inverse n-th power of $|{\cal D}|$ 
instead of the volume element $d^{n}x$. The Diximier trace of an operator 
$O$ is roughly speaking the logarithmic divergence of the ordinary trace:

$$ Tr_{\omega} O = \lim_{n\to\infty} \frac{\lambda _{1} + \dots +\lambda _{n}}
{log \ n}\ ,$$
where $\lambda _{i}$ is the i-th proper value of $O$. See [34-40] for details.
One can generalize the notions of covariant derivative ($\nabla $), 
connection ($A$) and curvature ($F$) forms so that "standard" properties are 
conserved:

$$\nabla = d + A \ \ , \ \ \ \ F= \nabla ^{2} = dA + A^{2}\ ,$$
where $A\in \Omega ^{1}_{\cal D}$ is the algebra of one forms defined with 
respect to $d$. Fiber bundles became projective modules on ${\cal A}$ 
in this language. The $n$-dimensional Yang-Mills 
fermionic action is given by the formula 
[35-38]

$$ {\cal L}\left( A, \psi, {\cal D}\right) = Tr_{\omega}\left( F^{2}\mid 
{\cal D}\mid ^{-n} \right) + <\psi \mid {\cal D}+A\mid \psi > \ ,$$
where $<\mid >$ denotes the inner product in the Hilbert space. 
For ${\cal A}= C^{\infty}(M)$ and ${\cal D}$ being the Dirac operator we 
recover the ordinary riemannian geometry of the spin manifold $M$. Physicists 
have learned from the noncommutative geometry that one can describe 
fundamental interactions by specifying the  Hilbert space of 
fermionic states and a representation of an $C^*$ algebra in this Hilbert 
space. If one takes $${\cal A}= C^{\infty}(M, {\bf C})\oplus  
C^{\infty}(M, {\bf H})\oplus M_{3\times 3} (C^{\infty}
(M, {\bf C}))\ ,\eqno(***)$$ 
the known fermionic states to span the Hilbert space and the generalized 
Dirac operator 
with the Kobayashi-Maskawa mass matrix as ${\cal D}$ one gets the standard 
model lagrangian [35-36]. The structure of the "world algebra" 
[36] (***) and the analysis given in the previous 
sections allow us to conclude that the 
spacetime structure is uniquely determined in the class of ${\bf R}$-compact 
spaces by fundamental interactions 
of fermions (gravitation is hidden in the metric tensor that "enters" the 
Dirac operator [35,40]) as the result of the properties of 
$C^{\infty}(M,{\bf C})$ and 
$C^{\infty}(M,{\bf H})$. The knowledge of $C^{\infty}(M)$ is sufficient for the 
construction of the manifold $M$ but the Higgs mechanism to be at work requires 
that $M$ be multiplied by some discrete space [34-40]. This means that 
we may not know the structure of the spacetime with  satisfactory precision 
but nevertheless fundamental interactions determine it in a quite unique way.
It should be noted here that if others rings would appear in (***) then 
this conclusion may not be true (for example, grand unified models can be 
less determinative than the "low energy approximation" [40]). 
Of course, it is still possible that the $C^{*}$ 
algebra ${\cal A}$ that describes 
correctly fundamental interactions do not correspond to any topological space. 
This would mean that spacetime can only approximately be described as a 
topological space, say, defined by some subalgebra of ${\cal A}$ or that 
fundamental interactions does not determine it uniquely. It should be stressed 
here that matter fields (fermions) and their interactions are 
essential in the process determining the 
spacetime structure. The pure gauge sector is insufficient because two 
$E$-compact spaces $X$ and $Y$ are homeomorphic if and only if the categories 
of all modules over $C(X,E)$ and $C(Y,E)$ are equivalent. The noncommutative 
geometry formalism even suggest that fermions define the spacetime via the 
Dirac operator at least on the theoretical level.\\ 

\ \ \ Let us now return to the smooth case. By using the heat kernel method 
[41] we can express the Yang-Mills action in the form [36, 40]:

$${\cal L}_{YM}\left( F\right) \sim \lim _{t\to 0}\frac{tr\left( F^{2}
exp\left( -t{\cal D}^{2}\right) \right)}{tr\left( exp \left(-t{\cal D}^{2}
\right) \right) } \ .$$
Suppose that we have a one parameter ($z$) family of differential structures 
and the corresponding family of Dirac operators ${\cal D}(z)$. The Duhamels's 
formula [40]
$$\partial _{z} \left( e^{-t{\cal D}^{2} \left( z\right) }\right) =
\int _{0}^{t} e^{-\left( t-s\right) \triangle \left( z\right) } \partial _{z} 
\left( {\cal D}^{2} \left( z\right) \right) 
e^{-s \triangle \left( z\right) }ds \ ,$$
where $\triangle $ is the scalar Laplacian, can be used to calculate the 
possible variation of ${\cal L}_{YM} (F)$ with respect to $z$.
Unfortunately our present knowledge of exoticness is to poor for 
performing such calculations. 
For an operator $K$ with a smooth kernel we have the following 
asymptotic formula [40]:
$$ tr \left( Ke^{-t{\cal D}^{2}} \right)  \sim tr \left( K\right) + \sum 
^{\infty} _{i=1} t^{i}a_{i}\ .$$
So if $F^{2}$ is smooth with respect to all differential structures (e.g. has 
compact support [27]) then the possible effects of exoticness are 
negligible. This means that we are unlikely to discover exoticness by 
performing "local" experiments involving gauge interactions. (Brans proved 
that exoticness can be localized in arbitrary small spatial region but they 
should cause extremely strong gravitational effects to be detectable.) 
If we consider only matter (fermions) coupled to gravity then the action can 
be expressed in terms of the coefficients of the heat kernel expansion of the 
Dirac Laplacian, ${\cal D}^{2}$ [40]. In this case we may be able 
to determine the differential structure only if the Dirac operator specifies 
it uniquely [26, 31].
In general case the possible physical effect of exotic smoothness is still 
an open problem. 

\section{Conclusions} 

\ \ \  We have analysed the problem of determining the spacetime structure. 
We should be able to determine the spacetime  in the class of 
${\bf R}$-compact spaces. We have to find a phenomenon that cannot be 
described  in terms of the algebra $C(M)$ to reject the assumption of 
${\bf R}$-compactness. If we are using only topological methods we will not 
be able to construct the topological model $M$ of the spacetime uniquely. An 
unbounded observable is necessary  to prove noncompactness of spacetime. 
In the general case, we will be able to construct only the Stone-\^ Cech 
compactification of the space in question. 
The existence  of a differential structure on $M$ allows for 
the identification 
of $M$ with the set of maximal ideals of $C^{\infty}(M)$, although we 
anticipate that the determination of the differential structure may 
be problematic. Connes' 
construction of the standard model lagrangian imply that fundamental 
interactions determine the model of spacetime in the class ${\bf R}$-compact 
spaces although more general models may not. Matter fields are essential for 
defining and determining the spacetime properties. If we are not able to 
determine $C(M, {\bf R})$ or $C(M, {\bf Q})$ then our knowledge of the 
spacetime structure is substantially limited. If this is the case we have 
a bigger class of spaces "at our disposal" and we are more free in making 
assumptions about the spacetime. \\ 

\ \ \ {\bf Acknowledgment:} I greatly enjoyed the hospitality 
 extended
to me during a stay at the Physics Department at the University of
Wisconsin-Madison. This work was supported in part by the grant 
{\bf KBN-PB 659/P03/95/08} and by the {\bf II Joint M. Sk\l odowska-Curie 
USA-Poland Fund grant MEN-NSF-93-145}.
\newpage
\begin {center} 
{\bf References}
\end{center}

\newcounter{bban}

\begin{list}
{[\arabic{bban}.]}{\usecounter{bban}\setlength{\rightmargin}
{\leftmargin}}
\item R. Sikorski, Introduction to differential geometry (PWN, Warsaw, 1972) 
(in Polish); P. Multarzy\' nski and M. Heller, Found. of Phys. 20 (1990) 1005
\item L. Gillman and M. Jerison, Rings of Continuous Functions  
(Springer Verlag, Berlin, 1986)
\item E. M. Vechtomov, Itogi Nauki i Tekhniki, ser Algebra, Topologia, 
Geometria (in Russian), vol. 28 (1990) 3
\item E. M. Vechtomov, Itogi Nauki i Tekhniki, ser. Algebra, Topologia, 
Geometria (in Russian), vol. 29 (1990) 119
\item E. Hewitt, Trans. Amer. Math. Soc. 64 (1948) 45
\item A. P. Balachandran et al, Syracuse Univ. preprint SU-4240-621 (1996) 
(hep-th );  R. D. Sorkin, Int. J. Theor. Phys. 30 (1991) 323; J. Kijowski, 
Rep. Math. Phys 11 (1977) 97
\item A. V. Arhangelskii, Topologiceskye prostranstva funkcii (Moscow 
Univ. Press, Moscow, 1989)
\item C. Isham, preprint gr-qc/960769
\item Chew Kim-Peu, Bull. Acad. Pol. Sci. ser. Math. Astron. Phys. 19 
(1971) 485
\item S. Mr\' owka, Acta Math. Acad. Sci. Hung 21 (1970) 239
\item I. M. Gelfand and A. N. Kolgomorov, Dokl. Acad. Nauk SSSR 22 (139) 11
\item E. S. Thomas,  Trans. Amer. Math. Soc. 126 (1967) 244
\item J. V. Whittaker, Ann. Math. 62 (1963) 74
\item P. Bankstone, J. Pure Appl. Algebra 97 (1994) 221
\item A. Ya. Helemskii, Banachovy i polynormirovannye algebry (Nauka, 
Novosibirsk, 1989)
\item R. V. Gamkrelidze, A. A. Argacev and S. A. Vahrameev, Itogi Nauki 
i Tekhniki, ser. Sovr. Probl. Matematiki (in Russian), vol. 35 (1989) 3
\item K. R. Goodearl, J. London Math. Soc. 19 (1977) 348
\item A. Prastaro, Rep. Math. Phys. 30 (1991) 273
\item N. M. J. Woodhouse, Geometric quantization (Oxford Univ. Press, New 
York, 1991)
\item M. Freedman, J. Diff. Geom. {\bf 17}, 357 (1982)
\item S. K. Donaldson, J. Diff. Geom. {\bf 18}, 279 (1983)
\item R. E. Gompf, J. Diff. Geom. {\bf 18}, 317 (1983)
\item S. DeMichelis and M. Freedman, J. Diff. Geom. {\bf 35}, 
219 (1992)
\item R. E. Gompf, J. Diff. Geom. {\bf 37}, 199 (1993)
\item C. H. Brans, Class. Quantum Grav. {\bf 11}, 1785 (1994); 
J. Math. Phys. 35 (1995) 5494
\item J. S\l adkowski, Acta Phys. Pol. {\bf B27}, 1649 (1996)
\item J. S\l adkowski, Int. J. Theor. Phys. 35 (1996) 2075
\item D. Freed and K. Uhlenbeck, Instantons and four-manifolds 
(Springer-Verlag, New York, 1984) 
\item Z. Bizaca and J. Etnyre, preprint dg-qa/9604007
\item H. Donnelly, Bull. London Math. Soc. {\bf 7}, 147 (1975)
\item M. Kreck and S. Stolz, Ann. Math. {\bf 127}, 373 (1988);
S. Stolz, Invent. Math. 94 (1998) 147
\item A. Wipf and S. D\" urr, Nucl. Phys. B443 (1995) 201
\item G. Schwarz and J. \' Sniatycki, Commun. Math. Phys 168 (1995) 441
\item A. Connes, Publ. Math. IHES {\bf 62} (1983) 44 
\item A. Connes, Noncommutative geometry (Academic Press, London, 1994)
\item J. G. V\'arilly and J. M. Garcia-Bond\'ia, J. Geom. Phys. 
{\bf 12}, 223 (1993); C. Martin, J. G. V\'arilly and J. M. Garcia-Bond\'ia, 
preprint hep-th/9605001 
\item A. Connes, in The interface of mathematics and physics
(Claredon, Oxford, 1990) eds . D. Quillen, G. Segal and S. Tsou.
\item A. Connes and J. Lott, Nucl. Phys. {\bf B} Proc. Suppl.
{\bf 18B}, 29 (1990)
\item J. S\l adkowski, Int. J. Theor. Phys. 33 (1994) 2381
\item A. H. Chamsedine, J. Fr\" olich and O. Grandjean, J. Math. Phys. 
36 (1995) 6255; A. H. Chamsedine, G. Felder and J. Fr\" olich, Nucl. Phys. 
B395 (1993) 672
\item N. Berline, E. Getzler and M. Vergne, Heat kernels and Dirac operators 
(Springer Verlag, Berlin, 1991)
\item P. B. Gilkey, Invariance, the heat equation and the Atiyah-Singer index 
theorem (Publish or Perish, Washington, 1984)
\item R. Ma\' nka and J. S\l adkowski, Phys. Lett. B224 (1989) 97 and 
Acta Phys. Pol. B21 (1990) 509

\item T. Kopf, preprint gr-qc/9609050
\end{list}

\end{document}